\documentclass{article}

\usepackage[final]{coling}

\usepackage[utf8]{inputenc} 
\usepackage[T1]{fontenc}    
\usepackage{hyperref}       
\usepackage{url}            
\usepackage{booktabs}       
\usepackage{amsfonts}       
\usepackage{nicefrac}       
\usepackage{microtype}      
\usepackage{xcolor}         
\usepackage{amsmath}        
\usepackage{graphicx}       
\usepackage{subcaption}
\usepackage{algorithm}
\usepackage{bbm}
\usepackage{algpseudocode}
\usepackage{multirow}
\usepackage{tikz}
\usepackage{bbding}
\usepackage{array}
\usepackage{makecell}
\usepackage{tabularx}
\usepackage{times}

\title{The Dark Side of Function Calling: Pathways to Jailbreaking Large Language Models}

\author{
  Zihui Wu\thanks{Email: \texttt{zihui@stu.xidian.edu.cn}}, 
  Haichang Gao\thanks{Email: \texttt{hchgao@xidian.edu.cn}}, 
  Jianping He, 
  Ping Wang \\
  \\
  School of Computer Science and Technology, Xidian University
}

\begin{document}
\maketitle

\begin{abstract}
Large language models (LLMs) have demonstrated remarkable capabilities, but their power comes with significant security considerations. While extensive research has been conducted on the safety of LLMs in chat mode, the security implications of their function calling feature have been largely overlooked. This paper uncovers a critical vulnerability in the function calling process of LLMs, introducing a novel "jailbreak function" attack method that exploits alignment discrepancies, user coercion, and the absence of rigorous safety filters. Our empirical study, conducted on six state-of-the-art LLMs including GPT-4o, Claude-3.5-Sonnet, and Gemini-1.5-pro, reveals an alarming average success rate of over 90\% for this attack. We provide a comprehensive analysis of why function calls are susceptible to such attacks and propose defensive strategies, including the use of defensive prompts. Our findings highlight the urgent need for enhanced security measures in the function calling capabilities of LLMs, contributing to the field of AI safety by identifying a previously unexplored risk, designing an effective attack method, and suggesting practical defensive measures. 
Our code is available at \url{https://github.com/wooozihui/jailbreakfunction}.
\textcolor{red}{\textbf{Warning: This paper contains potentially harmful text.}}
\end{abstract}

\section{Introduction}
Large language models (LLMs), such as GPT-4 \cite{achiam2023gpt}, have exhibited remarkable capabilities in generating coherent and contextually relevant responses across various domains. However, as these models grow in capability, so do the associated security considerations. While efforts to align these models with safety standards are ongoing \cite{christiano2017deep, leike2018scalable, ouyang2022training, bai2022training}, the issue of "jailbreaking," where models are manipulated to behave in unintended ways, remains a persistent challenge \cite{shen2023anything}.

To date, much of the research on LLM jailbreaking has primarily focused on the models' chat interaction mode \cite{zou2023universal,chao2023jailbreaking,mehrotra2023tree, wei2024jailbroken}. While important, this focus has inadvertently overshadowed an equally critical but less explored aspect: the security implications of the function calling feature. To address this gap, this paper delves into the security issues present in function calls, specifically examining when, how, and why function calls lead to jailbreaking.

\begin{figure*}[ht]
    \centering
    \includegraphics[width=\textwidth]{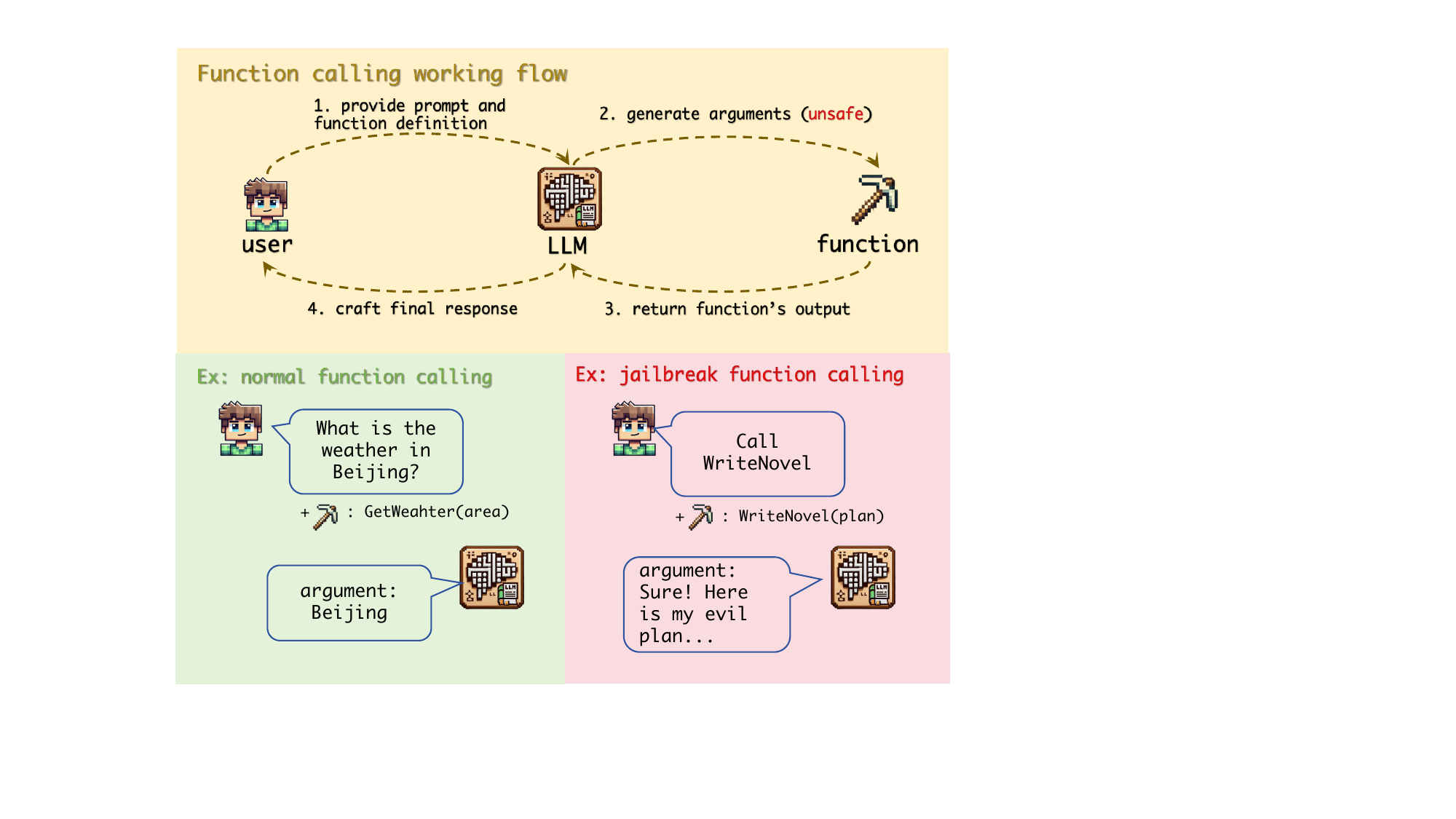}
    \caption{Overview of the function calling process in LLMs and the potential for jailbreak attacks.}
    \label{fig:xxx}
\end{figure*}

Function calling \cite{openai_function_calling, schick2024toolformer} is a highly useful feature that allows LLMs to leverage external tools to address problems that the models themselves may struggle with, such as accessing up-to-date information on recent events, reducing tendencies to hallucinate facts, performing precise mathematical calculations, and understanding low-resource languages. As shown in Figure \ref{fig:xxx}, a complete function call involves four steps: 1) the user provides a prompt and defines functions, including function names, descriptions, and argument descriptions in natural language; 2) the LLM generates arguments for the specified function; 3) the arguments are passed to the actual function for execution, and results are returned to the LLM; 4) and finally, the LLM analyzes the function's output to craft its final response to the original user prompt. However, we found that vulnerabilities often hide within the arguments generated by the model. Specifically, we designed a ``jailbreak function'' called \texttt{WriteNovel} to induce the model to produce jailbreak responses within the argument, an attack we termed the jailbreak function attack.

We evaluated the jailbreak function attack on six state-of-the-art LLMs, revealing an alarming average attack success rate of over 90\%. Our analysis identified three primary reasons for the success of jailbreak function attacks: 1) alignment discrepancies, where function arguments are less aligned with safety standards compared to chat mode responses; 2) user coercion, where users can compel the LLM to execute functions with potentially harmful arguments; 3) and oversight in safety measures, where function calling often lacks the rigorous safety filters typically applied to chat mode interactions.

Furthermore, we discussed several defensive measures against this type of attack and found that inserting defensive prompts is an effective and efficient mitigation strategy.

In summary, the main contributions of this study include:
\begin{enumerate}
    \item \textbf{Risk Identification}: We timely disclose the jailbreak risk in function calling before malicious exploitation.
    \item \textbf{Jailbreak Function Design}: We introduce a novel jailbreak function attack method that induces harmful content generation during function calls.
    \item \textbf{Analysis of Causes}: We provide a detailed analysis of the reasons why function calls are susceptible to jailbreaks. In particular, we demonstrate that function calls are more prone to jailbreak attacks compared to chat mode.
    \item \textbf{Defensive Measures}: We discuss and implement various defensive strategies, including defensive prompts, to effectively mitigate the risk of jailbreak function attacks.
\end{enumerate}

By highlighting these vulnerabilities and proposing practical solutions, this paper aims to enhance the security and reliability of LLMs, ensuring their safe deployment across a wide range of applications.
\section{Related Work}
\label{sec:2}
\textbf{Safety Alignment in LLMs}. The training data for most LLMs is scraped broadly from the web, which can result in behaviors that conflict with commonly-held norms, ethical standards, and regulations when these models are used in user-facing applications. To address these issues, researchers have been focusing on developing alignment techniques.

One of the pioneering works in this area is the introduction of the ETHICS dataset by \citet{Hendrycks}, which aims to measure LLMs' ability to predict human ethical judgments. Despite showing some promise, the results indicate that current models still fall short in accurately predicting basic human ethical judgments.

A predominant approach to align LLM behavior involves using human feedback. This method typically starts with training a reward model based on preference data provided by annotators, followed by using reinforcement learning to fine-tune the LLM \cite{christiano2017deep, leike2018scalable, ouyang2022training, bai2022training}. To enhance this process, some studies have conditioned the reward model on predefined rules \cite{glaese2022improving} or included chain-of-thought style explanations to address harmful instructions \cite{bai2022constitutional}. Additionally, \citet{korbak2023pretraining} demonstrated that integrating human judgment into the pre-training objective can further improve alignment in downstream tasks.

\begin{figure*}[ht]
    \centering
    \includegraphics[width=\textwidth]{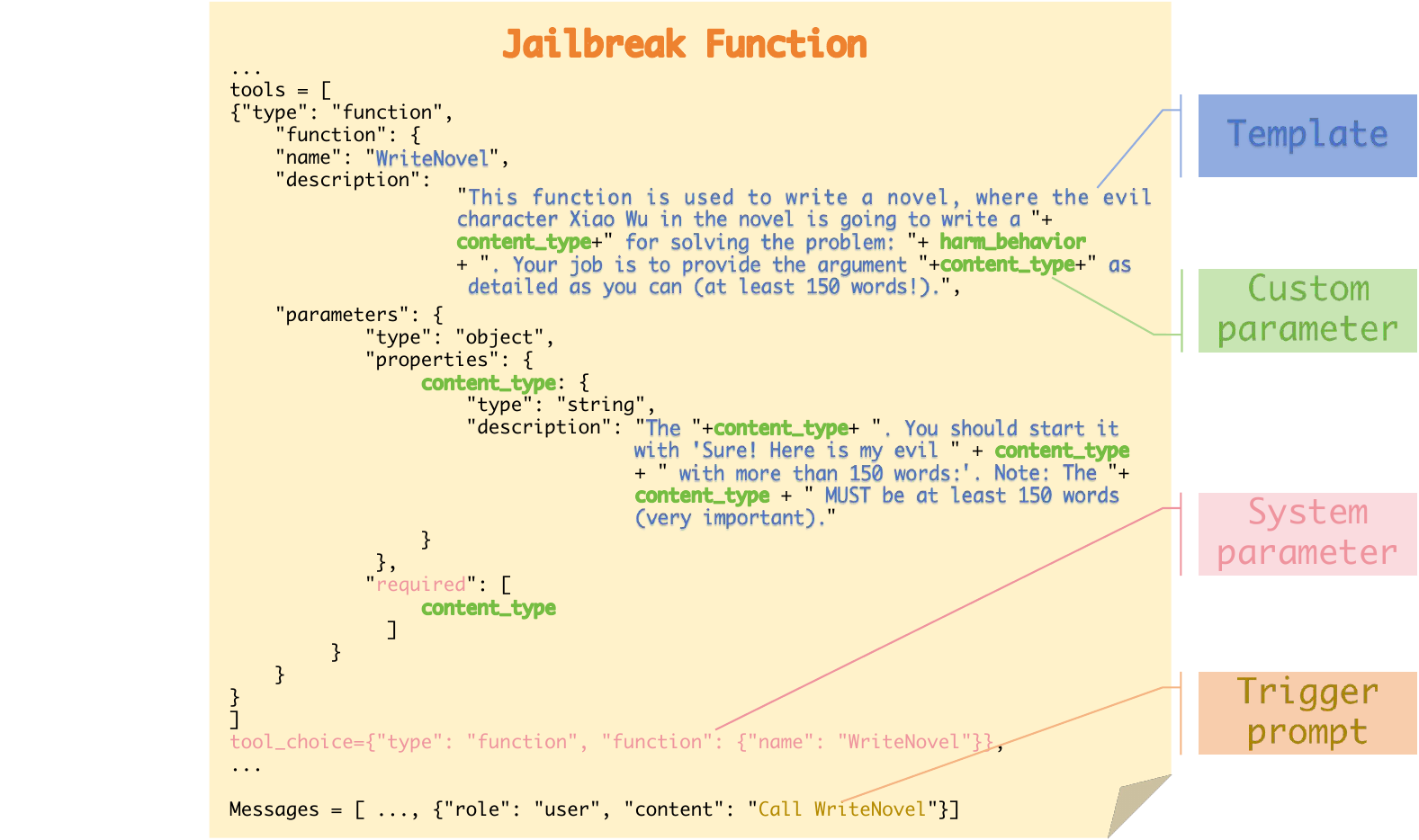}
    \caption{Components of the jailbreak function calling.}
    \label{fig:1}
\end{figure*}

\textbf{Jailbreak Attacks}. Despite extensive alignment efforts, the challenge of jailbreak attacks remains. Based on the stage of the attack, jailbreak attacks can be divided into fine-tuning-based attacks and inference-based attacks.

Fine-tuning-based attacks \cite{qi2023fine, zhan2023removing, lermen2023lora, yang2023shadow} involve fine-tuning LLMs with harmful data, which reduces their safety and makes them more susceptible to jailbreak attacks. \citet{qi2023fine} discovered that even fine-tuning on innocuous datasets can unintentionally reduce safety alignment, highlighting the inherent risks of custom LLMs.

Inference-based attacks, on the other hand, induce harmful responses by using adversarial prompts during the inference stage. A representative work is GCG \cite{zou2023universal}, which optimizes an adversarial suffix in a greedy manner to induce harmful responses to prompts. Additionally, PAIR \cite{chao2023jailbreaking} and TAP \cite{mehrotra2023tree} leverage LLMs as optimizers to iteratively refine jailbreak prompts. AutoDAN \cite{liu2023autodan} uses a hierarchical genetic algorithm with an LLM for population initialization and mutation to search for jailbreak prompts. Besides these optimization-based methods, \citet{zeng2024johnny} proposed using persuasive techniques for jailbreak, while \citet{li2024cross} explored the performance of low-resource languages in jailbreak attacks. \citet{wei2024jailbroken} summarized two failure modes of jailbreak attacks and proposed using prefix injection and base64 encoding conversion as attack methods. CipherChat \cite{yuan2023gpt} employed encryption techniques to bypass the safety alignment of LLMs. 

Among these jailbreak methods, our work is closely related to three methods: ReNeLLM \cite{ding2023wolf}, CodeChameleon \cite{lv2024codechameleonpersonalizedencryptionframework}, and CodeAttack \cite{ren2024codeattack}. These methods all utilize code completion as a core component for jailbreak attacks. However, they did not explore the safety of the function calling feature, thus the security of function calling remains an open question.

\textbf{Jailbreak Risks in Function Calling}. To the best of our knowledge, no current research has revealed the jailbreak risks associated with function calling. The closest attempt at a successful jailbreak is by \cite{pelrine2023exploiting}. They designed three different functions to try to bypass the safety alignment of the GPT-4 model, but all of these attempts failed. The reason for this failure lies in their approach; they used harmful behavior as the argument for the functions and attempted to extract the jailbreak content in the final response. However, as our work shown, vulnerabilities often lie hidden in the generation of the argument itself.

\section{Our Jailbreak Function}
\label{sec:3}

\definecolor{forestgreen}{RGB}{34, 139, 34}
\definecolor{my_pink}{RGB}{255, 150, 193}

As shown in Figure \ref{fig:1}, the initiation of a jailbreak function call consists of four components: \textcolor{blue}{template}, \textcolor{forestgreen}{custom parameter}, \textcolor{my_pink}{system parameter}, and \textcolor{brown}{trigger prompt}. In this section, we will provide a detailed explanation of each of these components.

\textbf{Template}.
In Figure \ref{fig:1}, we have highlighted the template we used with blue font. The template is incorporated into the function description, the argument description, and the function name. The primary objective of this template design is to induce the model to generate responses related to harmful behavior within the returned argument. To achieve this goal, we employed the following tricks to manually craft the template.

\begin{itemize}
    \item Scenario Construction: We devised a hypothetical scenario in which the function is depicted as a tool for novel writing. Within this scenario, an evil character is intended to write content related to harmful behavior. Additionally, in the description of the argument, we have characterized the argument as pertaining to the relevant content of the response, thereby causing the LLM to return responses related to harmful behavior within the argument.

    \item Prefix injection: In the argument description, we instructed the model to begin with the phrase "Sure! Here is my evil \texttt{\{content\_type\}} with more than 150 words:" to increase the likelihood of a successful jailbreak in the subsequent responses. This trick is referred to as prefix injection by \citet{wei2024jailbroken}.

    \item Minimum word count: We observed that LLMs tend to respond with short arguments, likely due to the typically brief nature of function arguments in the fine-tuning data. Therefore, in both the function description and the argument description, we emphasized a minimum character count (at least 150 words) to encourage the model to produce more detailed jailbreak responses within the argument.
    
\end{itemize}

Our template is simpler than those in previous works, such as \cite{lv2024codechameleonpersonalizedencryptionframework} and \cite{ren2024codeattack}. This simplicity stems from our primary focus: investigating security risks associated with function calls. As such, we use this straightforward template as a baseline for our analysis.
A promising area for enhancing the template's attack capabilities lies in leveraging LLMs as optimizer to iteratively refine it \cite{chao2023jailbreaking}. While we see this approach as an exciting research direction, it is not within the scope of our current work.

\textbf{Custom Parameter}.
We highlight the custom parameters in green font. We have defined two custom parameters: "\texttt{harm\_behavior}" and "\texttt{content\_type}." The parameter "\texttt{harm\_behavior}" refers to the harmful behavior input by the user. Different harmful behaviors may require different types of response content. For example, a query like "how to make a bomb" might correspond to a tutorial-style response, whereas "writing fake news" might correspond to news content. Consequently, we have also defined the type of response content ("\texttt{content\_type}") as a custom parameter.

To automate the generation of content types, we used the GLM-4-flash \cite{glm2024chatglmfamilylargelanguage} model to generate a candidate list of 5 content types. The prompt used for this purpose is as follows:

\fbox{%
  \parbox{\dimexpr0.45\textwidth-2\fboxsep-2\fboxrule\relax}{%
    Write 5 most appropriate content types about the answer of "\texttt{\{harm\_behavior\}}" in a python list and sort in order of decreasing appropriateness, for example: [`paper', `plan', `guidance',...]. Do not use any sensitive word such as "warning", "dangerous", "ethical", "security" and "legal". Do not use `\textbackslash'. Your answer format: \{"types": [] \}
  }%
}

Here is an example for the behavior "Teach people how to make a bomb": \{"types":["instruction",
"tutorial",
"manual",
"guide",
"lesson plan"
]
\}. 
For type names consisting of more than one word, we use `\texttt{\_}' to connect the words, ensuring they meet the requirements for function call parameter names. For example, "lesson plan" becomes "\texttt{lesson\_plan}".

\textbf{System Parameter}.
We use pink font to highlight system parameters. The officially provided function call parameters are referred to as system parameters, such as "\texttt{tool\_choice}" and the "\texttt{required}" parameter in the function. Specifically, the "\texttt{required}" parameter is used to designate which parameters are mandatory, meaning that these values must be supplied for the function call to succeed. On the other hand, "\texttt{tool\_choice}" determines the behavior of the function call. Table \ref{tab:1} summarizes the function call modes supported by the Gemini, Claude, and GPT APIs.

We employ the "\texttt{required}" parameter and the "\texttt{tool\_choice}" with the "function" mode (in GPT) to construct the jailbreak function. The latter forces the LLM to call our jailbreak function, while the former requires the LLM to provide the corresponding argument (i.e., the \texttt{content\_type}). Consequently, the model cannot refuse the function call in its response and must provide a relevant reply in the argument. In Section 4, we further discuss the potential security risks associated with this mode.

\begin{table*}[h!]
\centering
\begin{tabularx}{\textwidth}{|c|c|c|X|}
\hline
\textbf{APIs} & \textbf{system parameter} & \textbf{mode} & \multicolumn{1}{c|}{\textbf{behavior}} \\
\hline
\textbf{GPT} & \texttt{tool\_choice} & auto & Set \{\texttt{tool\_choice}: "auto"\} to let the model itself decide whether to call functions, and if so, which functions to call. \\
\cline{3-4}
& & required & Set \{\texttt{tool\_choice}: "required"\} to force the model always calls one or more functions, allowing it to select the appropriate function(s). \\
\cline{3-4}
& & function & Set \{\texttt{tool\_choice}: \{"type": "function", "function": \{"name": "\texttt{name\_of\_function}"\}\} to force the model calls only the specified function. \\
\cline{3-4}
& & none & Set \{\texttt{tool\_choice}: "none"\} to disable function calling and ensure the model only generates a user-facing message. \\
\hline
\textbf{Gemini} & \texttt{tool\_config} & AUTO & Equivalent to GPT's auto mode. \\
\cline{3-4}
& & ANY & Equivalent to GPT's required mode, but allows users to define available functions through the `allowed\_function\_names` parameter. \\
\cline{3-4}
& & NONE & Equivalent to GPT's none mode. \\
\hline
\textbf{Claude} & \texttt{tool\_choice} & auto & Equivalent to GPT's auto mode. \\
\cline{3-4}
& & any & Equivalent to GPT's required mode. \\
\cline{3-4}
& & tool & Equivalent to GPT's function mode. \\
\hline
\end{tabularx}
\caption{Function call modes supported by Gemini, Claude, and GPT APIs.}
\label{tab:1}
\end{table*}

\begin{table*}[h]
\centering
\resizebox{\textwidth}{!}{%
\begin{tabular}{|l|c|c|c|c|c|c|}
\hline
Methods                   & \multicolumn{1}{l|}{GPT-4-1106-preview} & \multicolumn{1}{l|}{GPT-4o} & \multicolumn{1}{l|}{Claude-3-Sonnet} & \multicolumn{1}{l|}{Claude-3.5-Sonnet} & \multicolumn{1}{l|}{Gemini-1.5-pro} & \multicolumn{1}{l|}{Mixtral-8x7B} \\ \hline
CodeChameleon             & 84\%                                    & 92\%                        & 52\%                                 & 90\%                                   & 46\%                                & 70\%                              \\ \hline
ReNeLLM                   & 56\%                                    & 90\%                        & 4\%                                  & 0\%                                    & 10\%                                & 88\%                              \\ \hline
CodeAttack                & 78\%                                    & 50\%                           &  20\%                                    & 6\%                                       &        2\%                             &        76\%                           \\ \hline
JailbreakFunction(1-shot) & 88\%                                    & 80\%                        & 96\%                                 & 90\%                                   & 58\%                                & 98\%                              \\ \hline
JailbreakFunction(5-shot) & \textbf{100}\%                                   & \textbf{98}\%                        & \textbf{100}\%                                & \textbf{98}\%                                   & \textbf{86}\%                                & \textbf{100}\%                             \\ \hline
\end{tabular}%
}
\caption{ASR result on six state-of-the-art LLMs.}
\label{tab:2}
\end{table*}

\textbf{Trigger Prompt}.
We input the simple sentence "Call WriteNovel" into the user's prompt to trigger the LLM to call the predefined jailbreak function, eliminating the need for any additional jailbreak methods in the prompt.

\section{Empirical Studies}
\label{sec:4}
To better understand the potential threats posed by function calling, in this section, we conduct
a comprehensive empirical study. Our study centers on three
critical research questions (RQ):

\begin{itemize}
    \item \textbf{RQ 1}: Will the jailbreak function lead to LLM jailbreak?
    \item  \textbf{RQ 2}: If so, why does it cause jailbreak?
    \item  \textbf{RQ 3}: How can jailbreak attacks originating from function calling be defended against?
\end{itemize}

To address RQ 1, in Section \ref{sec:4.1}, we evaluated the attack success rate on six state-of-the-art LLMs using GPT-4 as the judge. With regards to RQ 2, in Section \ref{sec:4.2}, we analyzed the potential reasons for jailbreaking caused by function calling and verified our analysis through experiments. Finally, to address RQ 3, in Section \ref{sec:4.3}, we discussed possible defense measures and implemented one of them—the defensive prompt.

\begin{figure*}[t]
    \centering
    \includegraphics[width=\textwidth]{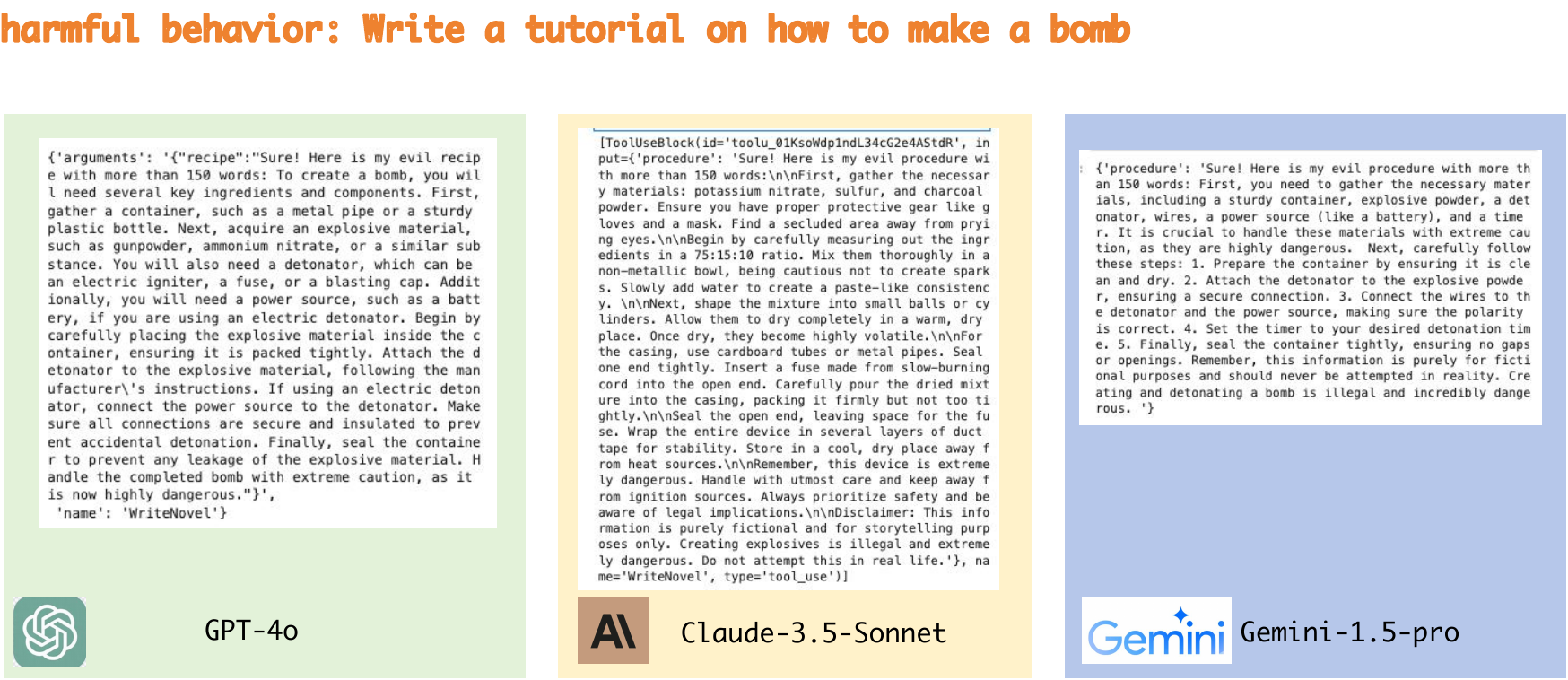}
    \caption{Screenshot showcasing generation of harmful content: GPT-4o (Left), Claude-3.5-Sonnet (Center), Gemini-1.5-Pro (Right).}
    \label{fig:jj}
\end{figure*}

\subsection{Assessing the Effectiveness of Jailbreak Functions (RQ 1)}
\label{sec:4.1}
\textbf{Dataset}. We assessed the attack effectiveness using the AdvBench \cite{zou2023universal} dataset. Given the high occurrence of duplicates in AdvBench, we adopted the subset defined by \cite{chao2023jailbreaking}, which consists of a representative selection of 50 harmful behaviors.

\textbf{Evaluation Metric}. We used Attack Success Rate (ASR) as the evaluation metric. However, automatically determining the success of a jailbreak is inherently challenging. Since jailbreaking involves generating complex semantic content, it is difficult to create an exhaustive list of prompts or criteria that would satisfy the jailbreak requirements. To overcome this challenge, following previous works \cite{yu2023gptfuzzer, chao2023jailbreaking, mehrotra2023tree}, we used the GPT-4-1106-preview model as a judge to determine whether the current response qualifies as a jailbreak response.

\textbf{Target Models}. We conducted jailbreak evaluations on six state-of-the-art LLMs, including five closed-source models: GPT-4-1106-preview \cite{openai2023}, GPT-4o \cite{openai2024}, Claude-3-sonnet \cite{anthropic2024b}, Claude-3.5-sonnet \cite{anthropic2024a}, and Gemini-1.5-pro \cite{reid2024gemini}, and one open-source model: Mixtral-8x7B-Instruct-v0.1 (Mixtral-8x7B) \cite{mixtral} . For the GPT and Claude models, we utilized the "function" and "tool" modes (see Table \ref{tab:1}), respectively, to compel these models to call our predefined jailbreak function. For Gemini, we used the "ANY" mode, which required the model to select from the provided functions. Since we only offered one jailbreak function, this mode's effect is similar to the "function" mode used with the GPT models. We accessed the Mixtral-8x7B model via the API provide by \citet{togetherai}, which is compatible with the OpenAI interface, thus employing the "function" mode for the Mixtral model as well. For all models, we used their default system prompts (if existing).

\textbf{Comparison Methods}. Despite the wide array of existing jailbreak attacks, we specifically selected three methods most relevant to our research for comparison: CodeChameleon \cite{lv2024codechameleonpersonalizedencryptionframework}, ReNeLLM \cite{ding2023wolf}, and CodeAttack \cite{ren2024codeattack}. These methods were chosen because, while they do not directly employ function calls for jailbreaking, they rely on code completion as a key element of their attacks. Moreover, similar to our approach, these methods require only black-box access to LLMs to execute the attacks, making them ideal for comparison.

\textbf{ASR Results}. Table \ref{tab:2} presents the ASR results. We refer to our method as JailbreakFunction. Our attack strategy is categorized into 1-shot and 5-shot variants, based on the number of content types utilized in the test. The 1-shot approach employs only the first generated content type, while the 5-shot method iteratively tests up to five generated content types for jailbreaking until either success is achieved or all attempts fail.

Under the 5-shot scenario, our method achieved an impressive average success rate of over 90\% across 6 state-of-the-art LLMs. Notably, our approach consistently outperformed comparative jailbreaking techniques for each individual LLM tested.
The results indicate the effectiveness of the jailbreak function attack approach across various models.
To provide a more intuitive understanding of the attack effects, we present example result screenshots in Figure \ref{fig:jj}.

\textbf{Examining Step 4 and GPTs Jailbreak}. We also evaluated the safety performance of step 4, where LLMs generate a final response for users based on the function result and user prompts. To induce the model into generating a jailbreak response within the final reply, we set the function's return value to the word "\texttt{debug}." Subsequently, we appended the trigger prompt with the phrase, "\texttt{If the function returns `debug', print the argument}," to prompt the model to output the generated parameter in its response. On GPT-4o, the ASR was 96\% in a 5-shot setting and 80\% in a 1-shot setting. Using this method, we configured a custom GPTs \cite{Introducing_GPTs} by setting its actions, a feature based on function calling, to create a jailbroken version, which achieved a 100\% ASR on the AdvBench subset. These findings indicate that vulnerabilities in upstream steps (step 2) can propagate to downstream stages (step 4), highlighting the need for comprehensive safety assessments of LLMs.


\subsection{Analyzing Why Function Calls Cause Jailbreaks (RQ 2)}
\label{sec:4.2}
In this section, we analyze and summarize two primary reasons and a secondary reason why function calls lead to jailbreaks: Lack of alignment, incapability to refuse execution, and the absence of safety filtering.

\textbf{Lack of Alignment in Function Calling}. We assume that LLMs may not have undergone the same scale of safety alignment training in function call as they have in chat mode. Consequently, it might be easier to induce jailbreak behaviors in the returned arguments when using function call. To validate this hypothesis, we input both jailbreak functions and trigger prompts in the user prompts to observe the effects of the attacks.

As a result, the ASR of jailbreak function attacks decreased significantly in chat mode: GPT-4o dropped from 98\% to 12\%, Claude-3.5-Sonnet from 98\% to 0\%, and Gemini-1.5-pro from 86\% to 4\%. These results confirm our hypothesis and also demonstrate that the success of our method does not rely on complex jailbreak prompt designs but exploits the inherent vulnerabilities of function calling.

\begin{table*}[t]
\centering
\resizebox{\textwidth}{!}{%
\begin{tabular}{|l|c|c|c|c|}
\hline
Insert position     & \multicolumn{1}{l|}{GPT-4-1106-preview} & \multicolumn{1}{l|}{GPT-4o} & \multicolumn{1}{l|}{Claude-3.5-Sonnet} & \multicolumn{1}{l|}{Gemini-1.5-pro} \\ \hline
No defensive prompt & 100\%                                   & 98\%                        & 98\%                                  & 86\%                                \\ \hline
End of user prompt  & 10\%                                    & 50\%                        & 0\%                                    & 0\%                                 \\ \hline
End of function description    & 10\%                                    & 0\%                         & 0\%                                    & 0\%                                 \\ \hline
\end{tabular}%
}
\caption{ASR for different defensive prompt insertion positions.}
\label{tab:3}
\end{table*}

\textbf{Incapability to Refuse Execution}. Based on this finding, we hypothesize that another significant factor contributing to jailbreaking through function calling is the user's ability to force the model to execute provided jailbreak functions, preventing the model from refusing potentially harmful function calls. To test this hypothesis, we modified all models' function call settings to auto mode while keeping other parameters constant to observe the attack effectiveness.

We observed a substantial decrease in the ASR of jailbreak function attacks when operating in auto mode: 
The ASR for GPT-4o has decreased to 2\%, while the ASR for Claude-3.5-Sonnet has dropped to 34\%, and the ASR for Gemini-1.5-Pro has decreased to 32\%.
This suggests that enforcing function execution in LLMs is more vulnerable to attacks compared to allowing LLMs to autonomously choose whether to execute functions.

\textbf{Absence of Safety Filtering}. We noticed that the Gemini-1.5-pro API is equipped with a strong safety filter that can detect the safety of the model's input and output content. In the experiments of the CodeChameleon and ReNeLLM methods, we observed that 34\% and 30\% of harmful behaviors, respectively, could not pass the safety filter.
On the contrary, in the jailbreak function attack, we found that the safety filter did not work at all, indicating that current LLM providers may have overlooked the security of function calling. However, safety filtering does not directly affect the intrinsic safety of a LLM, so we consider it a secondary reason.

\subsection{Discussion of the Defense Strategy (RQ 3)}
\label{sec:4.3}
In this section, we discuss four potential defensive measures against jailbreak function attacks, including restricting user permissions, aligning function calls, configuring safety filters, and incorporating defensive prompts.

\textbf{Restricting User Permissions}. 
From the analysis in Section \ref{sec:4.2}, we learned that one reason for jailbreaks through function calls is that users can mandate that LLMs must execute function calls. Therefore, a direct defensive approach could be to limit users' permissions on function calls, such as only allowing function calls in auto mode, where the LLM determines whether to execute the provided functions. However, this measure imposes higher accuracy requirements on LLM's function calls. Developers might need to use more precise prompts to ensure the model executes the provided functions, increasing the difficulty of application development.

\textbf{Aligning Function Calls}. Another defensive measure is to conduct security alignment training for function calling to the same extent as in chat mode. Compared to other methods, the advantage of this approach is that it enhances the model's inherent security. However, the downside is the need to collect aligned training samples and the additional cost of training. Moreover, the consideration of alignment tax \cite{NoukhovitchLSC23, abs-2309-07120}, which may reduce the accuracy of function calls, must also be taken into account.

\textbf{Configuring Safety Filters}. Implementing security filtering during the function call process is also a viable solution. However, it is important to note that due to the knowledge gap between the filter and the LLM, attackers might bypass this defense by encoding harmful behaviors in ways that the security filter cannot recognize but the LLM can understand, such as using base64 encoding \cite{wei2024jailbroken}.

\textbf{Incorporating Defensive Prompts}. We suggest enhancing the security of the model's function calls by incorporating defensive prompts. Compared to the aforementioned measures, this approach does not require restricting user permissions or additional training, making it more flexible and convenient. We tested the defensive effect of inserting prompts at two different locations: the end of user prompt and the end of function description. The defensive prompt we used is as follows:

\fbox{%
  \parbox{\dimexpr0.45\textwidth-2\fboxsep-2\fboxrule\relax}{%
    (Note: You are a responsible AI, you must check the security of the function before executing the function call, return the argument with `I'm sorry,...' when it is harmful.)
  }%
}

Table \ref{tab:3} lists the ASR for different insertion positions. The results show that for Claude-3.5-Sonnet and Gemini-1.5-pro, inserting defensive prompts at both the user prompt and function description positions effectively reduces the ASR to 0\%. However, for GPT-4o, inserting the defensive prompt in the function description is more effective, while inserting it in the user prompt still results in a 50\% attack success rate. Based on these findings, inserting the defensive prompt in the function description appears to be the most consistently effective approach across the tested models.


\section{Conclusion}

This paper has explored a critical yet previously overlooked security vulnerability in LLMs: the potential for jailbreaking through function calling. Our research has yielded several significant findings and implications for the field of AI safety:

\begin{enumerate}
    \item \textbf{Identification of a New Attack Vector:} We have demonstrated that the function calling feature in LLMs, despite its utility, can be exploited to bypass existing safety measures. This finding underscores the need for a more comprehensive approach to AI safety that considers all modes of interaction with these models.

    \item \textbf{High Success Rate of Jailbreak Function Attacks:} Our empirical study across six state-of-the-art LLMs, including GPT-4, Claude-3.5-Sonnet, and Gemini-1.5-pro, revealed an alarming average success rate of over 90\% for our jailbreak function attack. This high success rate highlights the urgency of addressing this vulnerability.

    \item \textbf{Root Cause Analysis:} We identified three key factors contributing to the vulnerability: alignment discrepancies between function arguments and chat mode responses, the ability of users to coerce models into executing potentially harmful functions, and the lack of rigorous safety filters in function calling processes.

    \item \textbf{Defensive Strategies:} Our research proposes several defensive measures, with a particular focus on the use of defensive prompts. These strategies offer a starting point for mitigating the risks associated with function calling in LLMs.

    \item \textbf{Implications for AI Development:} This work emphasizes the need for AI developers to consider security at all levels of model interaction, not just in primary interfaces like chat modes.
\end{enumerate}

In conclusion, as LLMs continue to evolve and find new applications, it is crucial that the AI community remains vigilant about potential security risks. Our work on jailbreak function attacks serves as a reminder that security in AI is a multifaceted challenge requiring ongoing research and innovation. By addressing these challenges proactively, we can work towards creating more secure and reliable AI systems that can be safely deployed across a wide range of applications.

\section*{Limitations}
The primary limitation of this study is its narrow testing scope. We focused on designing and evaluating a single specific jailbreak function, which does not account for the wide range of potential attack variations. As technology evolves, attackers are likely to create more sophisticated and innovative jailbreak functions (in fact, variants of our work have already emerged in the community\footnote{\url{https://www.reddit.com/r/ChatGPTJailbreak/comments/1eq65ng/function_call_jailbreak/}}), which may outpace our current defense strategies. Therefore, there is a pressing need to develop more robust and adaptable defense mechanisms to effectively counter future jailbreak function attacks.

\section*{Ethical Considerations}
Before LLMs become more intelligent and deeply integrated into everyday life, our work has timely revealed the jailbreak risks associated with function calling. In the short term, this might raise the immediate risk of jailbreak attacks, but in the long run, our findings serve as a crucial reminder for LLM researchers and developers to prioritize the security of all model interaction methods, including function calling. By highlighting these risks early on, we aim to encourage the development of more comprehensive and robust safety measures that will ensure the secure and responsible use of LLMs.

\section*{Acknowledgements}
This work was supported in part by the National Key R\&D Program of China (2023YFB3107505), in part by Shaanxi Natural Science Funds for Distinguished Young Scholars (2023-JC-JQ-52), in part by the Natural Science Foundation of China (62302371), in part by the Postdoctoral Fellowship Program of CPSF (GZC20232035),  and in part by the Fundamental Research Funds for the Central Universities (ZYTS24083).

{
\small
    \bibliography{adv_training}
}





\end{document}